\documentclass{iau}
\usepackage{graphicx}
\usepackage{natbib}
\usepackage{txfonts}
\usepackage[svgnames]{xcolor}
\makeatletter
\def\ps@plain{\let\@mkboth\@gobbletwo
     \def\@oddhead{\@journal}
     \let\@evenhead\@empty
     \let\@evenhead\@journal
     \def\@oddfoot{\reset@font{\vbox to 0pt{\vskip 50pt
     \parbox{13cm}{\raggedright\textcolor{DarkGreen}{\footnotesize\bf To
     appear in: ``Supernova environmental impacts'', eds A.\ Ray \&
     R.\ McCray, IAU Symposium No. 296 (Cambridge University Press),
     in press, 2013.}}\vss}}\hskip-7cm\thepage\hss}
     \let\@evenfoot\@oddfoot}
\def\@oddfoot{\vbox to 0pt{\vskip 50pt
\parbox{13cm}{\raggedright\textcolor{DarkGreen}{\footnotesize\bf To
appear in: ``Supernova environmental impacts'', eds A.\ Ray \&
R.\ McCray, IAU Symposium No. 296 (Cambridge University Press),
in press, 2013.}}\vss}}
\let\@evenfoot=\@oddfoot
\makeatother
\title[IC443 at 150~MHz]{A combined GMRT/CLFST image\\of IC443 at 150~MHz}

\author[Mitra et al.]{D.~Mitra$^1$, D.~A.~Green$^2$ and A.~Pramesh~Rao$^1$}

\affiliation{$^1$National Centre for Radio Astrophysics, Tata
Institute of Fundamental Research, Pune University Campus,
Post Bag 3, Ganeshkhind Pune 411007, India\\
email: {\tt dmitra@ncra.tifr.res.in}, {\tt
pramesh@ncra.tifr.res.in}\\[\affilskip]
$^2$Cavendish Laboratory, 19 J.~J.~Thomson Ave., Cambridge, CB3
0HE, U.K.\\email: {\tt dag@mrao.cam.ac.uk}}

\pubyear{2013}
\volume{296}
\pagerange{\pageref{firstpage}--\pageref{lastpage}}
\jname{Supernova environmental impacts}
\editors{A.\ Ray \& R.\ McCray eds.}
\begin{document}
\maketitle

\label{firstpage}

\begin{abstract}
IC443 is a relatively large Galactic ($\approx 45$~arcmin) SNR with a high
radio surface brightness. It has fine scale structure down to arcsec scales,
and so is difficult to image on all angular scales with a single instrument.
Here observations of IC443 at 151~MHz made with both the GMRT and the CLFST are
combined to give a composite image of IC443 on all scales from $>45$~arcmin
down to $\approx 20$~arcsec.
\end{abstract}

\firstsection

\begin{table}
\caption{Parameters of the GMRT and the CLFST.\label{t:telescopes}}
\centering
\begin{tabular}{rcc}\hline
                     & GMRT                   & CLFST                     \\\hline
number of antennas   & 30                     & 60                        \\
antenna type         & 45-m dish              & $4 \times 10$-element yagi\\
number of baselines  & 435                    & 776                       \\
longest baseline     & $\sim 25$~km           & $\sim 4.6$~km             \\
shortest baseline    & $\sim 100$~m           & $\sim 12$~m               \\
array layout         & 14 in central `square' & $\sim$ east--west         \\
                     & 16 in 3 arms           &                           \\
frequency            & 153~MHz                & 151.5~MHz                 \\
bandwidth            & 6~MHz                  & 0.8~MHz                   \\
primary beam         & $\sim 3^\circ$         & $\sim 17^\circ$           \\\hline
\end{tabular}
\end{table}

\section{Background}

IC443 ($=$G189.1$+$3.0) is a relatively bright SNR in the Galactic anti-centre,
where the Galactic background emission is relatively faint. It is $\approx
45$~arcmin in diameter, with brighter emission to the northeast, and fainter
emission (with a somewhat larger out radius) in the southwest. Structure is
seen at radio wavelengths down to scales of arcsec (e.g.\
\citealt{1989AJ.....98.1363D,1991AJ....102..224W}). Various observations show
that IC443 is interacting with a surrounding molecular cloud (e.g.\
\citealt{2007AJ....133...89R}), and radio spectral index studies reveal a
region of flatter spectrum emission in the east (see
\citealt{1986MNRAS.221..473G}).

\begin{figure}
\begin{tabular}{p{9.5cm}p{3.5cm}}
\centerline{\includegraphics[bb=73 178 530 672,angle=270,
  width=9.5cm,clip=]{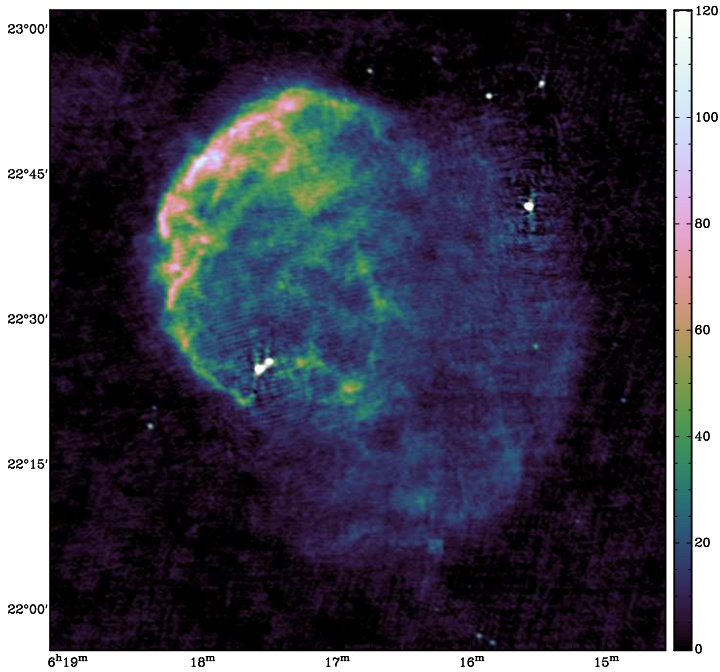}} &
\vspace*{3cm}
\caption{Combined GMRT plus CLFST image of IC443 with a resolution of $24''
\times 19''$~arcsec$^2$ (at PA of $61^\circ$). The scale is $0$ to $120$
mJy~beam$^{-1}$ (using the \textsc{`cubehelix'} colour scheme of
\citealt{2011BASI...39..289G}). The peak emission is $\approx
2.1$~Jy~beam$^{-1}$, from the background source near the NE edge of the
remnant, and the noise is $\approx 2.0$~mJy~beam$^{-1}$.\label{f:combined}}
\end{tabular}
\end{figure}

\section{Observations and Results}

We have combined observations made of IC443 at 151~MHz from two telescopes (see
Table~\ref{t:telescopes}), in order to cover a wide range of angular scales.
The Giant Metrewave Radio Telescope (GMRT) -- see \citet{2002IAUS..199..439R}
-- is a synthesis telescope that provides baselines up to $\sim 25$~km, but
lacks good $uv$-plane coverage on baselines less than a few hundred metres. The
GMRT observations of IC443 miss about 20\% of the total expect flux density of
the SNR ($\approx 280$~Jy at 151 MHz, \citealt{1986MNRAS.221..473G}), due to
the missing short baselines. The Cambridge Low-Frequency Synthesis Telescope
(CLFST) -- see \citet{1990MNRAS.244..233R} -- was an (approximately) E--W
synthesis telescope that provides good coverage of the $uv$-plane for the small
baselines missed by the GMRT. The smoothed 151-MHz image of IC443, from
observations made in 1983/84, as used in \citet{1986MNRAS.221..473G}, with a
resolution of $5'\!\!.4 \times 2'\!\!.1$~arcmin$^2$ (NS$\times$EW) covers
baselines up to $\sim 1$~km at all position angles. Figure~\ref{f:combined}
shows the combined GMRT plus CLFST image of IC443 at 151~MHz. This includes
emission on a wide range of scales from $\approx 20$~arcsec to 45~arcmin. This
was made using the {\tt IMERG} task in {\sc AIPS} which takes the
larger/smaller scale structure from the CLFST/GMRT images respectively,
gradually merging the contributions on the intermediate scales in both images.

\begin{acknowledgments}
We thank staff of the Mullard Radio Astronomy Observatory and the National
Centre for Radio Astrophysics.
\end{acknowledgments}


\label{lastpage}


\begin{thebibliography}{}

\bibitem[Dickel et al.(1989)Dickel et al.]{1989AJ.....98.1363D}
  Dickel J.~R., Williamson C.~E., Mufson S.~L., Wood C.~A., 1989,
  \textit{AJ}, 98, 1363

\bibitem[Green(1986)Green]{1986MNRAS.221..473G}
  Green D.~A., 1986, \textit{MNRAS}, 221, 473

\bibitem[Green(2011)Green]{2011BASI...39..289G}
  Green D.~A., 2011, \textit{BASI}, 39, 289

\bibitem[McGilchrist et al.(1990)McGilchrist et al.]{1990MNRAS.246..110M}
  McGilchrist M.~M., Baldwin J.~E., Riley J.~M., Titterington D.~J.,
  Waldram E.~M., Warner P.~J., 1990, \textit{MNRAS}, 246, 110

\bibitem[Pramesh Rao(2002)Pramesh Rao]{2002IAUS..199..439R}
  Pramesh Rao A., 2002, in: A.\ Pramesh Rao, G.\ Swarup \& Gopal-Krishna
  (eds), \textit{The Universe at Low Radio Frequencies}, Proc.\ IAU
  Symposium No.~199, (San Francisco: ASP), p.~439

\bibitem[Rees(1990)Rees]{1990MNRAS.244..233R}
  Rees N., 1990, \textit{MNRAS}, 244, 233

\bibitem[Rosado et al.(2007)Rosado, Arias
  \& Ambrocio-Cruz]{2007AJ....133...89R}
  Rosado M., Arias L., Ambrocio-Cruz P., 2007, \textit{AJ}, 133, 89

\bibitem[Wood et al.(1991)Wood, Mufson \& Dickel]{1991AJ....102..224W}
  Wood C.~A., Mufson S.~L., Dickel J.~R., 1991, \textit{AJ}, 102, 224

\end{thebibliography}
\end{document}